# Studying and Classification of the Most Significant Malicious Software

Dr. Wajeb GHARIBI,
Computer Science & Information Systems College, Jazan University,
Jazan, KSA.
gharibi@jazanu.edu.sa

*Abstract*—**As the cost of information processing and Internet accessibility falls, most organizations are becoming increasingly vulnerable to potential cyber threats which its rate has been dramatically increasing every year in recent times.**

**In this paper, we study, discuss and classify the most significant malicious software: viruses, Trojans, worms, adware and pornware which have made step forward in the science of Virology.**

## I. Introduction

Nowadays, there is a huge variety of cyber threats that can be quite dangerous not only for big companies but also for ordinary user, who can be a potential victim for cybercriminals when using unsafe system for entering confidential data, such as login, password, credit card numbers, etc. Among popular computer threats it is possible to distinguish several types of them depending on the means and ways they are realized. They are: malicious software (malware), DDoS attacks (Distributed Denial-of-Service), phishing, banking, exploiting vulnerabilities, botnets, threats for mobile phones, IP-communication threats, social networking threats and even spam. All of these threats try to violate one of the following criteria: confidentiality, integrity and accessibility. Lately, malicious software has turned into a big business. Cyber criminals became profitable organizations and able to perform any type of attack. An understanding of today's cyber threats is vital part for safe computing and ability to counteract the cyber invaders.

The rest of our paper is organized as follows: Section 2 demonstrates the theory of computer viruses. Section 3 proposes the history: from the first viruses till the last epidemics. Section 4 defines and classifies Malware. Conclusions have been made in Section5.

## II. Theory of Computer Viruses

The history begins in 1983, when American scientist Fred Cohen in the dissertational work devoted to research of self-reproducing computer programs for the first time has proposed the term 'computer virus' and later on published the article «Computer Viruses: theory and experiments»[1].

Len Eidelmen first coined the term 'virus' in connection with self-replicating computer programs. On November 10[th], 1983, at a seminar on computer safety at Lehigh University, this grandfather of modern computer virology, demonstrated a virus-like program on a VAX11/750 system[2].

Nevertheless, the idea for computer viruses actually appeared much earlier. Many consider the starting point to be the work of John Von Neumann in his studies on self-reproducing mathematical automata, famous in the 1940s. By 1951, Neumann had already proposed methods for demonstrating how to create such automata. In 1959, the British mathematician Lionel Penrose presented his view on automated self-replication in his Scientific American article 'Self-Reproducing Machines'. Unlike Neumann, Penrose described a simple two dimensional model of this structure which could be activated, multiply, mutate and attack. Shortly after Penrose's article appeared, Frederick G. Stahl reproduced this model in machine code on an IBM 650 [3]. It should be noted that these studies were never intended to providing a basis for the future development of computer viruses. On the contrary, these scientists were striving to perfect this world and make it more suitable for human life. Afterwards, these works established the foundation for many later studies such as robotics and artificial intelligence.

## III. History: from the first viruses till the last epidemics

Sometime in the early 1970s, the Creeper virus was detected on ARPANET, a US military computer network which was the forerunner of the modern Internet. Written for the then-popular Tenex operating system, this program was able to gain access independently through a modem and copy itself to the remote system. As computers gained in popularity, more and more individuals started writing their own programs. Advances in telecommunications provided convenient channels for sharing programs through open-access servers such as BBS - the Bulletin Board System.

*Elk Cloner* virus infected the boot sector for Apple II computers and spread by infecting the operating system, stored on floppy disks.

*Brain* was the first global IBM-compatible virus epidemic, which infected the boot sector, and was able to spread practically worldwide within a few months. It was written by a 19-year-old Pakistani programmer, Basit Farooq Alvi, and his brother Amjad, and included a text string containing their names, address and telephone number. Interestingly enough, *Brain* was the first 'stealth virus.'; when one attempts to read the detected infected sector, the virus would display the original, uninfected data.

Another such hoax was released by Robert Morris about a virus spreading over networks and changing port and drive configurations. According to the warning, the alleged virus infected 300,000 computers in the Dakotas in under 12 minutes. November 1988: a network epidemic caused by the Morris Worm. The virus infected over 600 computer systems in the US (including the NASA research center) and almost brought some to a complete standstill. In order to multiply, the Morris Worm exploited vulnerability in UNIX operating systems on VAX and Sun Microsystems platforms. As well as exploiting the UNIX vulnerability, the virus used several innovative methods to gain system access such as harvesting passwords. The overall losses caused by the 'Morris Worm' virus were estimated at US$96 million dollars - a significant sum at that time.

CodeRed, Nimda, Aliz and BadtransII were the Malicious programs that exploited vulnerabilities in applications and operating systems and caused serious epidemics in 2001. The large-scale epidemics caused by these worms changed the face of computer security and set trends for malware evolution for several years to come. Moreover, 2001 was also the year that instant messaging services, such as ICQ and MS Instant Messenger, were first used as channels for spreading malicious code [4].

Email worms, such as Klez and Lentin had already been popular prior to 2002. However, a new breed of email worms superseded the older versions: these new email worms spread by connecting directly to built-in SMTP servers on infected machines. Worms multiplying in other environments, such as LANs, P2P, IRC and so forth, disappeared almost entirely in this year. Though Klez caused the most serious outbreak during 2002, several other worms provided some stiff competition: Lentin and Tanatos (aka Bugbear).

In 2003 two global Internet attacks took place that could be called the biggest in the history of the Internet. The Internet Worm Slammer laid the foundation for the attacks, and used vulnerability in the MS SQL Server to spread. Slammer was the first classic fileless worm, which fully illustrated the capabilities of a flash-worm - capabilities which had been foreseen several years before. The worm attacked computers through ports 1433 and 1434 and on penetrating machines did not copy itself on any disk, but simply remained in computer memory. If we analyze the dynamics of the epidemic, we can assert that the worm originated in the Far East.

The second, more important epidemic was caused by the Lovesan Worm, which appeared in August 2003. The worm demonstrated just how vulnerable Windows is. Just as Slammer did, Lovesan exploited vulnerability in Windows in order to replicate itself. The difference was that Lovesan used a loophole in the RPC DCOM service working under Windows 2000/XP. This led to almost every Internet user being attacked by the worm.

In February 2004 appeared Bizex (also known as Exploit) - the first ICQ-Worm. The unauthorized distribution of ICQ message «http://www.jokeworld.biz/index.html:)) LOL» was used to spread widely. After installing into the system, Bizex closed running ICQ-client and connected to ICQ server with the data of the infected user and started up delivery to all contacts from the list. At the same time, there was the theft of confidential data - banking data, various user logins and passwords.

In the same in 2004 the so-called war of malware writers is occured. Several criminal gangs are known for worms Bagle, Mydoom and Netsky released new versions of their programs literally every hour. Each new program carries a regular message to the opposing faction, full of threats, Netsky even removed any found worms copies of Mydoom and Bagle.

Mail Worm Bagle was firstly detected on 18 January 2004. To spread, it used its own SMTP-client, the worm code are sent as an attachment with a random name and extension .exe. Delivery was made to addresses found on infected machine. Bagle also contained a built-backdoor- procedure that opened port 6777 to run commands and download any files.

Mydoom is primarily known by a massive 12-day DDoS-attack on the Web site of SCO company, which began on the first of February 2004. In response, leaders of SCO announced a reward of $250 thousand dollars for information on the author of the worm. To spread, Mydoom uses mail delivery through its own SMTP-client, as well as P2P-network (Kazaa)[5].

Sasser (May 2004) - struck more than 8 million computers, the loss from this worm are estimated in $979 millions. For penetration Sasser used a vulnerability in Service LSASS Microsoft Windows.

Cabir (June 2004) - the first network worm which propagates via Bluetooth and infects mobile phones running OS Symbian. The viruses for Pocket PC are appeared soon (August 2004), - classic virus Duts and Trojan Horse Brador. However, malicious software - is not only viruses and Trojans. This class is also included adware - programs that perform unauthorized displaying on-screen advertisment, and pornware - programs that self-initiate a connection to paid pornographic websites. Since 2004 indicated widespread use of viral technology to install adware/pornware on target computers. Next epidemic of network Worm Kido/Conficker/ Downadup (November 2008) - has struck more than 10 million computers, using vulnerability in service "Server" (MS08-067). The new variant Kido loaded at night from 8 to 9 of April, 2009 (Net-Worm. Win32. Kido.js). One more very dangerous threat 2008/09 became bootkit Backdoor.Win32.Sinowal. Bootkit 2009 is distributed through the cracked sites, porno resources and sites from which it is possible to download pirate software [6].

IV. MALWARE DEFINITION AND CLASSIFICATION

Computer virus definition is a complicated problem, because it's quite difficult to give an efficient virus definition by showing properties attributable to viruses only and not concerning the other program systems. Let us give the following definitions:

*Definition 4.1.* The opportunity of making duplicates (the copies could not match with the original) and theirs embedding into the computer network and/or files, system computer areas and other executive objects, is the most required computer virus property. Meanwhile the duplicates can be distributed.

The other problem associated with computer virus definition is misapprehension of a virus. So any malware could be a virus. This leads to confusion in terminology, which is complicated by the ability of modern antivirus programs to detect specified types of malware. That is why the association "malware-virus" is getting more settled. Hence, malicious programs could mean viruses.

*Definition 4.2.* Malware is a computer program or a portable code which aimed to damage the information, stored in computer network or hidden use of computer network resources, or the other impact, which interrupt normal operation of computer network.

Computer viruses, Trojans and Worms are the malware fundamental types. Every malware includes subclasses of malicious programs which named according those functions, which were described above.

Viruses can be classified according to distribution methods because the distinctive feature is the ability to proliferate within the computer. Distribution process could be divided into several stages:

- o Penetration into computer
- o Virus activation
- o Objects search for infection
- o Preparation of virus duplicates
- o Distribution of virus duplicates

It is necessary for virus activation that infected object gets a control. So viruses divided according to the objects types that can be infected:

*Boot viruses* – viruses which infect boot sectors of hard and removable disks. For example, malicious program Virus.Boot.Snow.a writes its code into HDD MBR or into floppy disc boot sectors.

*File viruses* – viruses that infect files. This group is divided to three subgroups, depending on the environment where the code is executing.

*Actually file viruses* – viruses that work directly with OS resources. For example, the virus: Virus.Win9x.CIH also known as "Chernobyl". It has little size (about 1 kb) this virus infects PE-files (Portable Executable) under Windows95/98 control in the way that size of the infected files is not changed.

*Macro viruses* – viruses that are created by using the macro command language and executable in the environment of any application. It is talked of Microsoft Office Macro in most cases. For example, Macro.Word97.

*Script viruses* – viruses which are executable in the certain command frame environment: firstly bat-files in the command frame DOS, nowadays VBS and JS – scripts in the command frame Windows Scripting Host (WSH). For example, Virus.VBS.Sling has been written by using VBScript (Visual Basic Script) language. Once launched it searches files with .VBS and .VBE extension and infects them. On June, 16 or July, 16 this virus deletes all files with .VBS and .VBE extension including itself.

*Definition 4.3.* Worm (net-worm) is a type of malicious programs which can distribute by network channels. Worms can run independently through security systems of automated and computer networks. They can create and distribute their duplicates which are not coinciding with the original and realize different harmful operations.

Worm Life Cycle can be divided to the following stages:

- o Penetration into computer
- o Activation
- o "Victims" search
- o Preparation of duplicates
- o Distribution of duplicates

The stages 1 and 5 are symmetric and defined by the used of protocols and applications. There is no difference between stage 4 and the stage in virus distribution process. This conception is applicable to worms which can be divided by the types of the used protocol [7]:

*Network worms* – worms that use Internet and local networks protocols. Usually, the worms of this type can be distributed by mistake in processing of base packets in TCP/IP by some applications.

*Email worms* – worms that are distributed by email messages.

*IRC worms* – worms that are spread by IRC (Internet Relay Chat) channels.

*P2P worms* – worms that are spread by P2P (peer-to-peer) networks.

*IM worms* – worms that are distributed by using instant messaging applications (IM, Instant Messenger – ICQ, MSN Messenger, AIM etc.)

*Definition 4.4.* Trojan (Trojan Horse) is such type of malicious programs which goal is harmful effect to computer network. Trojans have no the mechanism of its duplicates creation. Some Trojans can bypass computer network security system to penetrate and infect system. In general case Trojan gets into system with virus or worm by active intruder's acts or by heedless user's operations. Trojans have no distribution function and its life cycle is short – only three stages:

- o Penetration to computer
- o Activation
- o Performing malicious functions

Let us consider the Trojan classification by Kaspersky Lab grouped according to three types of information threat that may violate (Figure 1):

*Type 1. Confidentiality*

*Backdoor* – remote administration utilities that open infected machines to external control via a LAN or the Internet.

*PSW Trojan* – steals passwords from the system.

*Trojan-Spy* – includes a variety of spy programs and key loggers, all of which track and save user activity on the victim machine and then forward this information to the master.

*Trojan-GameThief* – steals the user information pertaining to online games.

*Trojan-Banker* – steals the user information pertaining to the banking system, the electronic money and plastic cards.

*Trojan-Mailfinder* – provides unauthorized collection of user email addresses with the subsequent transfer to the attacker [3].

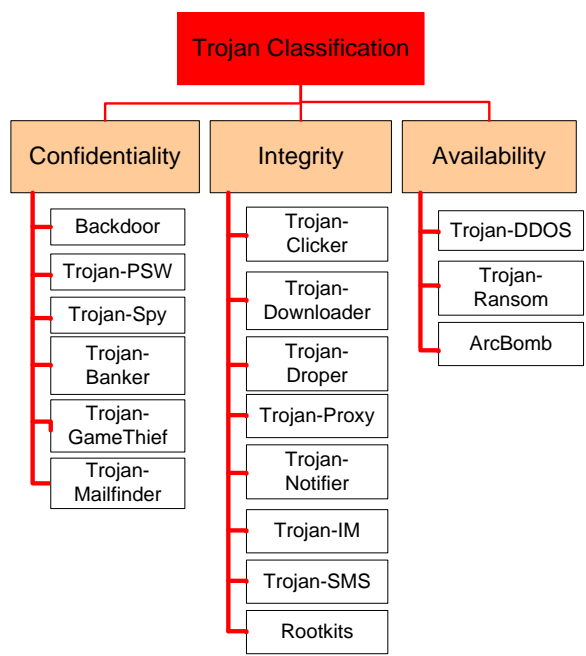

Fig.1. Trojan classification

*Type 2. Integrity*

*Trojan Clicker* – redirects victim machines to specified websites or other Internet resources.

*Trojan Downloader* – downloads and installs new malware or adware on the victim machine.

*Trojan Dropper* – used to install other malware on victim machines without the knowledge of the user.

*Trojan Proxy* – function as a proxy server and provide anonymous access to the Internet from victim machines.

*Trojan-Notifier* – inform the 'master' about an infected machine.

*Trojan-IM* – steals user's account (login and password) from the Internet-pager (e.g., ICQ, MSN Messenger, AOL Instant Messenger, Yahoo Pager, Skype, etc.)

*Trojan-SMS* – used for unauthorized sending SMS-messages from the compromised mobile devices to expensive paid numbers that are stored in the malware body.

*Rootkits* – a collection of programs used by a hacker to evade detection while trying to gain unauthorized access to a computer.

*Type 3. Availability*

*Trojan-DDoS* – performs an unauthorized DoS (Denial of Service) attack from infected computers to a computer-sacrifice with the specified address.

*Trojan-Ransom* - used for unauthorized data modification on victim's computer to make it impossible to work with it or block the normal functioning of the computer.

*ArcBomb* - archived files coded to sabotage the de-compressor when it attempts to open the infected archived file.

Other malware includes a range of programs that do not threaten computers directly, but are used to create viruses or Trojans, or used to carry out illegal activities such as DoS attacks and breaking into other computers [8].

## V. CONCLUSIONS

The given information cannot cover all variety of global information threats, however shows the most actual trends in area of cybercriminal. It is obvious the continuous improvement of "white" and "black" technologies in struggle of the anti-virus companies against army of hackers. In the triangle « a hacker - AV company - a user » the weak part still remains the user which should know the information about existed cyber threats to be able to use the installed protection system effectively.

## Brief Biography

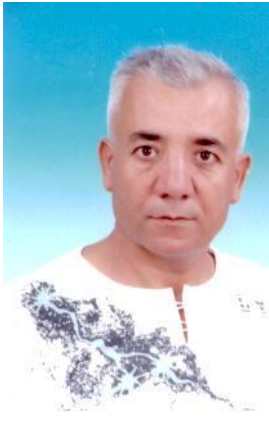

Wajeb Gharibi is an Associate Professor, Chairman, Department of Computer Networks, College of Computer Science & Information Systems, Jazan University, Jazan, Kingdom of Saudi Arabia.

He obtained his Ph. D degree in Informatics from Institute of Mathematics and Computer Science, Byelorussian Academy of Sciences, USSR in 1990.

Dr. Wajeb worked in Aleppo University, Syria (1990-1994, 1998-2001), Taiz University, Yemen (1995-1998), Saudi Arabia; King Khalid University (2001-2009), and Since October 2009, he has been teaching at the College of Computer Science & Information Systems, Jazan University, Jazan, Kingdom of Saudi Arabia.

His research interests include Information Security, Microelectronics; embedded systems, Design & analysis of network algorithms.. Combinatorial optimization, computational geometry, discrete convexity, operations research and data analysis.

He has published more than 53 research papers in national and international journals and conferences.

Dr. Wajeb Gharibi got many prizes:

- Proclamation of Great Minds of 21st Century; American Biographical Institute, USA (2008 and 2010)

- 2008 Man of the Year in Science Award; American Biographical Institute, USA

- 2000 Outstanding Intellectuals of the 21st Century Award, International Biographical Centre, Cambridge, England (2008)

- TWAS (Third World Academy of Sciences) Prize for the year 2001, Italy; (Supreme Council of Sciences, Syria).